# Self-Reported AI Usage for Learning in Computer Science Education: Relationships with Goal Orientation and Academic Help-Seeking

Piret Luik[†]
Institute of Computer Science
University of Tartu
Tartu, Estonia
piret.luik@ut.ee

Karin Naruskov
Institute of Education
University of Tartu
Tartu, Estonia
karin.naruskov@ut.ee

Karmen Kalk
Institute of Education
University of Tartu
Tartu, Estonia
karmen.kalk@ut.ee

Merle Taimalu
Institute of Education
University of Tartu
Tartu, Estonia
merle.taimalu@ut.ee

## ABSTRACT

Artificial intelligence (AI) is becoming an increasingly integral part of higher education, yet the factors shaping students' use of AI for learning remain insufficiently understood. This study examines how students' goal orientation and academic help-seeking behavior are associated with AI use in a computer science context, while also accounting for individual, behavioral, and contextual characteristics. Data were collected from 236 university students enrolled in a database course using a self-report survey. AI use was operationalized through two measures: self-reported frequency of use and the number of AI-supported learning activities. Hierarchical regression analyses were conducted to examine the relationships among the variables. The results indicate that help-seeking tendencies, particularly perceived help-seeking threat, were consistently associated with both more frequent reporting of AI use and self-reported engagement in a wider range of AI-supported activities. In contrast, the effects of goal orientation were more limited and less consistent across models. The results highlight the importance of considering help-seeking behavior when designing AI-supported learning environments in computer science education.

## 1 Introduction

The use of artificial intelligence (AI) has the potential to fundamentally transform teaching and learning practices in higher education [1, 2], including computer science (CS) education [3, 4]. Prior research suggests that CS students are familiar with AI tools and hold positive attitudes toward their use [5]. Generative AI tools have increasingly become a source of academic support for students and can be viewed as a contemporary form of help-seeking [6, 7, 8]. From this perspective, AI tools represent an additional source of academic support alongside instructors, peers, and learning materials.

However, despite the widespread availability of AI tools, students differ in how often they use them and the extent to which they rely on them for learning [9]. These differences may be related to students' motivational orientations, help-seeking tendencies, and various individual, behavioral, and contextual characteristics. However, evidence regarding these factors remains limited. Results regarding gender differences in AI use are mixed, particularly in CS education. Some studies suggest that male students use AI tools more frequently [10], whereas others report no substantial differences in overall AI use [11]. At the same time, little is known about the role of learning-related behaviors and course-related contextual characteristics, such as study mode or course requirements, in shaping students' AI use. Understanding these differences is important for explaining how students integrate AI into their learning processes and for identifying the characteristics associated with AI use as a source of learning support.

Because programming tasks are often complex and cognitively demanding [7, 12], understanding students' help-seeking behavior, including through AI tools, has become increasingly important. At

*Article Title Footnote needs to be captured as Title Note
[†]Author Footnote to be captured as Author Note

Preprint. Submitted to the ACM SIGCSE Technical Symposium 2027.

the same time, students differ in the motivational orientations that guide their learning behavior. Goal orientation and academic help-seeking are conceptually and empirically related constructs, with prior research showing that students' motivational orientations are closely linked to how they seek and use help in learning contexts [13, 14]. However, students' use of AI may also be shaped by individual characteristics, learning behaviors, and contextual features of the course environment. Identifying the characteristics associated with AI use may help educators better understand how different students use AI as a source of learning support.

The aim of this study is to examine how students' goal orientation and academic help-seeking behavior are associated with AI use for learning, and how these relationships are influenced by individual, behavioral, and contextual characteristics. Accordingly, the study addresses the following research question: How are students' goal orientation and academic help-seeking behavior, together with individual, behavioral, and contextual characteristics, associated with self-reported AI use for learning?

# 2 Theoretical background

## 2.1 Academic help-seeking

Academic help-seeking is a self-regulated learning strategy defined as the process of seeking assistance from individuals or other sources to achieve desired goals in learning [15]. Several components of help-seeking, such as students' goals, perceived threat, the source of help, and help-seeking avoidance, affect students' decisions about whether and how to seek help [16]. In academic contexts, help-seeking involves various sources. In recent years, the emergence of AI tools, particularly ChatGPT, has made it possible for students to access immediate and personalized forms of support without asking questions publicly [6, 7, 8], which is particularly valued by CS students who are more introverted or who experience fear related to asking for help from others [7].

Based on students' goals, help-seeking can be categorized as executive (or expedient) and instrumental (or adaptive). *Executive help-seeking* involves seeking direct answers or solutions, whereas instrumental help-seeking involves requesting support that enables independent problem solving, while *Instrumental help-seeking* means requesting just enough support to enable autonomous work [16]. *Avoidance of help-seeking* occurs when a student needs help but does not ask for it, for example, by skipping a question or writing any answer instead of seeking assistance [15]. *Help-seeking threat* refers to students' concern that asking for academic help might harm their self-esteem [17]. In large classes, however, students may perceive greater anonymity when seeking help, which can reduce some of the social risks associated with help-seeking [13].

The way students engage with these different forms of help-seeking can meaningfully influence their learning process and outcomes [18]. In CS education, help-seeking behaviors may differ from those observed in other disciplines because students often rely on online resources and self-directed problem solving [7]. CS students tend to start with convenient but less effective sources – such as online materials and peers – and turn to more formal, higher-utility options like instructors only if needed [18]. This pattern reflects how they balance convenience with effectiveness when seeking help. CS students' help-seeking may vary by task. It was found that CS students relied on traditional resources for technical help and task-specific guidance, while ChatGPT was used to explore ideas, gain multiple perspectives, and support creative or flexible problem-solving [7]. These findings suggest that AI use may be closely intertwined with CS students' help-seeking tendencies, making academic help-seeking a potentially crucial factor in understanding differences in AI use for learning.

## 2.2 Achievement Goal Orientation

Achievement goal orientation describes the purposes or goals that guide students' behavior in achievement situations and influence how they interpret, approach, and respond to learning tasks [19]. Early research distinguished between mastery and performance goals [19, 20]. *Mastery-oriented* students seek to develop competence and understanding, whereas *performance-oriented* students focus on demonstrating competence relative to others. Later models further differentiated performance goals into approach and avoidance orientations, resulting in the Trichotomous Model of Achievement Goal Orientation [21, 22].

Achievement goals are closely related to help-seeking behavior. Mastery goals have consistently been associated with instrumental help-seeking, whereas performance-oriented goals have been linked to more expedient forms of help-seeking [13, 14]. Mastery orientation has been found positively related to instrumental help-seeking and negatively related to help-seeking threat, executive help-seeking, and avoidance [8]. In CS education, students generally begin courses with relatively high mastery orientations, but these orientations may change over time and both mastery and performance goals declined during the semester, while mastery-avoidance increased [23].

Goal orientation has also been associated with AI use in learning. AI-supported learning has been linked to the development of mastery-oriented learning processes [24]. In addition, students with mastery and performance-approach orientations have been found to use AI less frequently than students with stronger avoidance-oriented goals [25]. Research in CS education further suggests that students with mastery goals may value AI not because it completes tasks for them, but because it can support learning and increase confidence during problem solving [26].

# 3 Methodology

## 3.1 Study Context

"Databases" is a mandatory core course in the computer science (CS) curriculum at the University of XXX. Owing to its relevance across multiple disciplines, the course is also attended by students from several other curricula, either as a compulsory or elective course. Consequently, students differ in their prior knowledge, learning goals, and motivation, providing an opportunity to examine how motivational, behavioral, and contextual characteristics are associated with AI usage for learning. Approximately 300 students enroll in the course each spring, making individualized support challenging and providing relevant context for examining AI as a source of learning support.

Practical sessions (approximately 20 students per group) are offered in two formats: on-campus and online. Students are free to choose their preferred study mode, creating different learning contexts within the same course. In addition, because the course serves different curricula as either a compulsory or elective subject, it provides an opportunity to examine contextual characteristics related to AI use.

### 3.2 Sample

In the spring 2025, 345 students were enrolled in the "Databases" course. Of these 236 students completed the voluntary online questionnaire and constituted the sample of the present study. The gender distribution of the sample was 155 (65.7%) males and 70 (33.5%) females. Two respondents did not mark their gender. More than half of the students (193 students) studied computer science, mathematics or statistics curricula. Among the respondents were also students from several other curricula (e.g. Economics and Business Administration, Biology, Physics, Medicine, English Language and Literature, Semiotics).

Age was not collected as part of the questionnaire to preserve participants' anonymity. Given the presence of small subgroups from certain curricula, the combination of demographic variables could have made individual participants identifiable. This decision was made in accordance with institutional ethical guidelines.

### 3.3 Data Collection

Data were collected using an anonymous online questionnaire administered during the spring semester of 2025. Participation was voluntary and anonymous, informed consent was obtained from all participants, and the study followed institutional ethical guidelines.

The questionnaire consisted of four parts. The first part measured goal-orientation according to Trichotomous Model of Achievement Goal Orientation based on Elliot and Dweck [19] main dimensions of the goal orientation. The items measuring academic help-seeking using Karabenick's [18] four behaviors formed the second part of the questionnaire. 5-point Likert scale (1 - totally disagree, …., 5 - totally agree) was used in the first two parts of the instrument.

Both scales were translated and adapted into Estonian and has previously been used in a similar higher education context with CS students. The reliability of both scales was examined in the present dataset. Previous validation analyses of the goal orientation scale in this context supported a two-factor structure consisting of *Mastery Approach* and *Performance Approach*; therefore, only these factors were included in the present study. Three items measured *Mastery Approach* (Cronbach's alpha (α) was 0.84 and McDonald's omega (ω) was 0.85) and three items measured *Performance Approach* (α=0.72 and ω=0.78) as goal-orientation factors. Academic help-seeking was measured by four factors: Instrumental Help-seeking (α=0.66 and ω=0.68), Expedient Help-seeking (α=0.67 and ω=0.68), Help-seeking Threat (α=0.83 and ω=0.84) and Help-seeking Avoidance (α=0.88 and ω=0.89).

The third part measured using AI in the learning on that course. At first, students had to answer the question 'How often did you use AI in learning on this course?' using a 5-point scale (1- never, 2 - rarely, 3 - sometimes, 4 - often, 5 - always). If their response was at least 2, they were shown a list of 13 activities, which were formulated by researchers considering the learning activities and tasks of the course (e.g., getting an overview of the theory topic, solving SQL tasks, generating teamwork, etc.). The activities were selected from the responses of open question in year 2024 where students who used AI in this course reported in which activities in the course they used AI. Students responded on a similar 5-point scale to indicate how often they used AI in this activity during the course.

The questionnaire ended with a section for collecting background data about the individual (gender, prior knowledge in SQL) and contextual characteristics (study mode, i.e., whether the student was in the online group or not; course requirement status, i.e., whether the course was compulsory, optional or elective for the student). In addition, two questions were used measuring students' learning behavior: how many hours on average did you spend learning (less than 2 hours, 2-4 hours, ... 8-10 hours, more than 10 hours) and when did you seek help if the problem arose (immediately upon reading the task, as soon as the first problem arose, if I could not find a solution from the materials, after a long time trying to solve it myself)?

### 3.4 Data Analysis

IBM SPSS Statistics 30.0 and Amos 30.0 were used for data analysis. AI use was operationalized using two measures: self-reported frequency of AI use and the number of AI-supported learning activities. The latter was calculated as an aggregate variable representing the breadth of AI use across different learning activities.

Hierarchical multiple regression analysis with the enter method was used to examine associations between AI use and the predictor variables. For both dependent variable (AI use frequency and number of AI-supported activities), two models were estimated. Model 1 included goal orientation and academic help-seeking factors, whereas Model 2 additionally included individual, behavioral, and contextual characteristics. Multicollinearity was assessed using variance inflation factor (VIF) values, with values below 10 indicating no serious multicollinearity issues [27]. The normality of residuals was evaluated using skewness and kurtosis values, with values between –1 and 1 considered indicative of acceptable normality [28].

## 4 Results

A hierarchical multiple regression analysis was conducted to examine the associations between the predictors and the self-reported frequency of AI use. In Model 1, which included goal orientation and academic help-seeking variables, the model explained 16.8% of the variance in AI usage frequency, $R^2 = .168$,

Adjusted $R^2$ = .130, F(6, 130) = 4.380, p < .001. In Model 2, individual, behavioral, and contextual variables were added (Table 1), increasing the explained variance to 30.8%, $R^2$ = .308, Adjusted $R^2$ = .241, F(12, 124) = 4.598, p < .001. This increase was statistically significant, $\Delta R^2$ = .140, $\Delta F$(6, 124) = 4.18, p < .001. The values of skewness (0.660) and kurtosis (0.458) indicate that the standardized residuals were approximately normally distributed.

**Table 1. Hierarchical regression analysis predicting self-reported frequency of AI use for learning**

| Predictor | Model 1 β | T | Model 2 β | T | VIF[a] |
|---|---|---|---|---|---|
| **Motivational variables** | | | | | |
| Mastery Approach | -0.275 | -2.849*** | -0.243 | -2.641** | 1.516 |
| Performance Approach | 0.042 | 0.461 | 0.034 | 0.373 | 1.459 |
| Instrumental Help-seeking | 0.315 | 3.385*** | 0.257 | 2.842** | 1.464 |
| Expedient Help-seeking | 0.122 | 1.387 | 0.171 | 1.951 | 1.372 |
| Help-seeking Threat | 0.214 | 2.207** | 0.220 | 2.230* | 1.751 |
| Help-seeking Avoidance | 0.016 | 0.167 | -0.077 | -0.758 | 1.847 |
| **Control variables (Model 2 only)** | | | | | |
| Gender | - | - | -0.044 | -0.514 | 1.332 |
| Course requirement status | - | - | 0.050 | 0.629 | 1.138 |
| Study mode of practicals | - | - | 0.138 | 1.782 | 1.080 |
| Time investment in the course | - | - | 0.335 | 4.291*** | 1.093 |
| Timing of help-seeking | - | - | -0.114 | -1.403 | 1.191 |
| Prior knowledge | - | - | 0.030 | 0.366 | 1.221 |

a- VIF values are shown for Model 2, * p<0.05, ** p<0.01, *** p<0.001

A similar hierarchical regression approach was applied to the self-reported number of AI-supported learning activities (Table 2). In the initial model, goal orientation and academic help-seeking variables accounted for 11.5% of the variance, $R^2$ = .115, Adjusted $R^2$ = .074, F(6, 127) = 2.761, p = .015. When individual, behavioral, and contextual variables were included, the explained variance increased to 24.6%, $R^2$ = .246, Adjusted $R^2$ = .171, F(12, 121) = 3.284, p < .001, with a significant change in model fit, $\Delta R^2$ = .130, $\Delta F$(6, 121) = 3.48, p = .003.

Compared to the model predicting frequency of AI use, the overall explained variance was lower, suggesting that the included factors may be more strongly related to how often students use AI rather than to the range of activities for which it is used. Despite the significant improvement in Model 2, the explained variance remained modest, indicating that additional factors may contribute to students' use of AI across different learning activities. Again, the skewness (0.324) and kurtosis (0.324) of the standardized residuals demonstrated that normality assumptions were met.

**Table 2. Hierarchical regression analysis predicting number of self-reported AI use activities for learning**

| Predictor | Model 1 β | T | Model 2 β | T | VIF[a] |
|---|---|---|---|---|---|
| **Motivational variables** | | | | | |
| Mastery Approach | -0.190 | -1.895 | -0.166 | -1.701 | 1.521 |
| Performance Approach | -0.025 | -0.255 | -0.019 | -0.193 | 1.479 |
| Instrumental Help-seeking | 0.214 | 2.204* | 0.158 | 1.661 | 1.459 |
| Expedient Help-seeking | 0.095 | 1.023 | 0.126 | 1.351 | 1.403 |
| Help-seeking Threat | 0.280 | 2.054* | 0.238 | 2.278* | 1.751 |
| Help-seeking Avoidance | 0.042 | 0.411 | -0.060 | -0.561 | 1.854 |
| **Control variables (Model 2 only)** | | | | | |
| Gender | - | - | 0.011 | 0.122 | 1.329 |
| Course requirement status | - | - | 0.039 | 0.459 | 1.147 |
| Study mode of practicals | - | - | 0.191 | 2.334* | 1.077 |
| Time investment in the course | - | - | 0.294 | 3.558*** | 1.095 |
| Timing of help-seeking | - | - | -0.144 | -1.668 | 1.190 |
| Prior knowledge | - | - | -0.017 | -0.195 | 1.226 |

a- VIF values are shown for Model 2, * p<0.05, ** p<0.01, *** p<0.001

Although instrumental help-seeking was significantly associated with self-reported number of AI use activities in Model 1, this effect did not remain significant in the full model. Given that instrumental help-seeking was only weakly to moderately correlated with other predictors (correlation with time investment 0.18 (p<0.05), with mastery approach 0.36 (p<0.01) and with expedient help-seeking –0.32 (p<0,01)), this suggests that its association with number of AI activities is relatively unstable and may depend on the inclusion of additional variables in the model.

## 5 Discussion

This study examined how students' goal orientation and academic help-seeking behavior are associated with AI use for learning, and how these relationships are influenced by individual, behavioral, and contextual factors. Considering the results across both models and outcome measures, the findings suggest that students' academic help-seeking tendencies play a more central role in explaining AI use in learning than goal orientation. In particular, perceiving help-seeking as threatening was consistently associated with both the frequency of AI use and the number of AI-supported activities, even when individual, behavioral, and contextual factors were taken into account. In contrast, the effects of goal orientation were more limited and less consistent across models. These results suggest that the use of AI as a source of learning support may be more strongly related to how students approach seeking help than to their underlying motivational goals, highlighting the importance of considering help-seeking behavior when examining students' engagement with AI tools.

When predicting the frequency of AI use, Model 1 showed that mastery approach was negatively associated with AI usage, whereas instrumental help-seeking and help-seeking threat were positively associated. After adding individual, behavioral, and contextual variables in Model 2, these associations remained statistically significant, and time investment in the course also emerged as a positive predictor.

For the number of AI-supported learning activities, Model 1 indicated that both instrumental help-seeking and help-seeking threat were significantly associated with AI use. However, in Model 2, only help-seeking threat remained statistically significant. This suggests that the association between instrumental help-seeking and the number of AI activities is not independent and may be explained by other factors included in the model, whereas help-seeking threat shows a more robust relationship. In addition, participation in online practicals and time investment in the course were positively associated with the number of AI activities.

The model explained more variance in AI usage frequency than in the number of AI-related activities, suggesting that students' motivational and behavioral characteristics may be more strongly related to how often they use AI rather than to the breadth of activities in which AI is used.

Among goal orientation factors, Mastery Approach was negatively associated with the self-reported frequency of AI use. Similarly, it was found in prior study [25], that students with a high mastery orientation were less likely to use ChatGPT. One possible explanation is that students who aim for deep understanding may limit their use of AI due to concerns about reliability, academic integrity, or reduced cognitive effort [4, 6]. However, in the present study, mastery approach was not associated with the self-reported number of AI-supported activities, suggesting that such students may restrict how often they use AI rather than the ways in which they use it. This interpretation is also consistent with previous findings that CS students with mastery-oriented goals still valued AI-supported learning and reported increased confidence when completing programming tasks independently [26].

Help-seeking Threat emerged as the most consistent predictor of AI use, being positively associated with both self-reported frequency and the number of activities. This finding supports the view of AI as a contemporary form of help-seeking, particularly for students who perceive interpersonal help-seeking as threatening. Previous research similarly indicates that AI tools can provide a non-judgmental and readily available form of assistance, reducing concerns about appearing less competent [6, 7, 8]. In contrast, help-seeking avoidance was not significantly associated with AI use, suggesting that students who generally avoid seeking help do not necessarily substitute human help with AI tools.

Instrumental Help-seeking was positively associated with AI use self-reported frequency but did not remain significant for the self-reported number of activities in the full model. This indicates that the students who extensively use AI tools for learning pointed out that AI provides different perspectives and additional sources of support beyond their instructors [4]. However, while students who actively seek hints and explanations may use AI more frequently, this tendency does not independently predict the breadth of AI use when other factors are considered. Together with the findings for help-seeking threat, this result suggests that AI may serve as a source of academic support that complements rather than replaces traditional forms of help-seeking [7].

Time investment in learning was positively associated with both self-reported AI usage frequency and the number of activities. This may indicate that students who spend more time studying encounter more challenges and therefore make greater use of AI tools as support. Alternatively, students who perceive AI as useful may integrate it more extensively into their learning processes, which in turn increases their overall engagement [2, 5]. At the same time, previous studies have shown that AI can improve efficiency, suggesting that its use may both increase and streamline learning efforts [6].

Participation in online practicals was positively associated with the self-reported number of AI-supported activities, which may reflect differences in access to immediate support. Students in online settings may rely more on AI tools when instructor or peer support is less readily available. This finding highlights the importance of contextual factors in shaping AI use.

Contrary to expectations, whether the course was compulsory or elective was not associated with AI use. This suggests that AI use may not simply reflect lower motivation or attempts to minimize effort in mandatory courses but may instead represent a learning strategy used by students regardless of their curricular reasons for taking the course. In addition, gender was not a significant predictor of AI use in either model. This finding contrasts with Luik [10] but is consistent with other studies reporting no substantial gender differences in AI use [11]. Taken together, these findings suggest that AI use may be more strongly related to students' learning behaviors and help-seeking tendencies than to demographic or curricular characteristics.

## 6 Conclusion

By conceptualizing AI as a potential source of academic support, the study contributes to understanding why students differ in their use of AI tools for learning. Although the study was conducted in a database course, the findings may be relevant to other CS courses where students engage in cognitively demanding problem-solving tasks and have access to AI tools. In such contexts, students' help-seeking tendencies and learning behaviors may similarly influence how AI is used as a source of learning support.

The findings suggest that students' help-seeking tendencies, particularly perceived help-seeking threat, are more consistently associated with AI use than goal orientation. Students who perceived help-seeking as threatening reported both more frequent use of AI and engagement in a wider range of AI-supported learning activities. These findings support the view of AI as a form of academic help-seeking and suggest that students' perceptions of help-seeking may be an important factor shaping how AI is used for learning in computing education.

This study has several limitations that should be considered when interpreting the findings. First, the sample consisted of students from a single course ("Databases"), which may limit the generalizability of the results to other CS contexts. Future studies could extend this work by examining AI use across multiple courses, disciplines, and institutions. In addition, the study relied on self-reported data, which may not fully reflect actual AI usage. Students may under- or overestimate their use of AI due to recall bias or social desirability.

Although the study included a range of individual, behavioral, and contextual variables, it did not capture all factors that may be relevant for understanding AI use in learning. Age was not collected in this study due to ethical considerations related to preserving participants' anonymity. Future research could incorporate these additional variables to provide a more comprehensive understanding of students' AI use.

These limitations and the rapid development of AI technologies should be considered, as changes in tools and usage practices may affect the applicability of the findings over time. Future research is therefore needed to examine how AI uses evolve and how it relates to learning outcomes in different contexts.

## ACKNOWLEDGMENTS

This work was sponsored by the Estonian Research Council grant "Developing human-centric digital solutions" (TEM-TA120).